\documentclass[conference]{IEEEtran}
\IEEEoverridecommandlockouts
\usepackage{cite}
\usepackage{amsmath,amssymb,amsfonts}
\usepackage{algorithmic}
\usepackage{graphicx}
\usepackage{textcomp}
\usepackage{xcolor}
\usepackage{color}
\usepackage{multirow}
\def\BibTeX{{\rm B\kern-.05em{\sc i\kern-.025em b}\kern-.08em
    T\kern-.1667em\lower.7ex\hbox{E}\kern-.125emX}}

\usepackage{comment}

\newcommand{\bi}[1]{\ensuremath{\boldsymbol{#1}}}   
\newcommand{\argmax}{\mathop{\rm arg~max}\limits} 
\newlength\savedwidth
\newcommand{\wcline}[1]{\noalign{\global\savedwidth\arrayrulewidth\global\arrayrulewidth 1.0pt} \cline{#1}
\noalign{\global\arrayrulewidth\savedwidth}}

\begin{document}

\title{Acoustic Scene Classification Using Multichannel Observation with Partially Missing Channels
}

\author{\IEEEauthorblockN{Keisuke Imoto}
\IEEEauthorblockA{Doshisha University, Japan \\
keisuke.imoto@ieee.org}
}

\maketitle
\begin{abstract}
Sounds recorded with smartphones or IoT devices often have partially unreliable observations caused by clipping, wind noise, and completely missing parts due to microphone failure and packet loss in data transmission over the network. In this paper, we investigate the impact of the partially missing channels on the performance of acoustic scene classification using multichannel audio recordings, especially for a distributed microphone array. Missing observations cause not only losses of time-frequency and spatial information on sound sources but also a mismatch between a trained model and evaluation data. We thus investigate how a missing channel affects the performance of acoustic scene classification in detail. We also propose simple data augmentation methods for scene classification using multichannel observations with partially missing channels and evaluate the scene classification performance using the data augmentation methods.
\end{abstract}


\section{Introduction}
Acoustic scene classification (ASC), which classifies sound recordings into the predefined class such as recording environments, places, and daily activities, is one of the core search problems in environmental sound analysis \cite{Virtanen_Springer2018_01,Imoto_AST2018_01,Heittola_DCASE2020_01}.
ASC has significant potential for various applications such as monitoring infants/elderly people \cite{Mesaros_DCASE2017c_01}, automatic surveillance \cite{Ntalampiras_ICASSP2009_01}, automatic life-logging \cite{Imoto_IEICE2016_01}, and media retrieval \cite{Jin_INTERSPEECH2012_01}.

Many methods for ASC utilizing spectral information have been proposed.
For instance, Eronen {\it et al.} \cite{Eronen_TASLP2006_01} and Mesaros {\it et al.} \cite{Mesaros_EUSIPCO2010_01} have proposed methods based on mel-frequency cepstral coefficients (MFCCs) and Gaussian mixture models (GMMs).
Valenti {\it et al.} \cite{Valenti_IJCNN2017_01}, Han {\it et al.} \cite{Han_DCASE2017_01}, and Jallet {\it et al.} \cite{Jallet_DCASE2017_01} have proposed methods using mel-spectrograms and a convolutional neural network (CNN).
Liping {\it et al.} \cite{Liping_DCASE2018_01}, Tanabe {\it et al.} \cite{Tanabe_DCASE2018_01}, and Raveh and Amar \cite{Raveh_DCASE2018_01} have proposed Xeption-based, VGG-based, and ResNet-based ASC methods, respectively .

More recently, environmental sound analysis utilizing spatial information, which is extracted from time differences or sound power ratios between channels, has also been studied \cite{Kwon_ISCS2009_01,Giannoulis_EUSIPCO2015_01,Imoto_TASLP2017_01,Nakadai_DCASE2018_01,Tanabe_DCASE2018_01,Imoto_IEICE2020_01}.
Conventional microphone array processing requires that microphones are synchronized between channels and/or microphone locations or array geometry is known.
However, spatial information based on accurate time differences or sound power ratios between channels cannot be extracted using a combination of unsynchronized distributed microphones such as smartphones, IoT devices, and surveillance cameras.
To utilize unsynchronized distributed microphones whose locations or array geometry is unknown for multichannel ASC, K\"{u}rby \textit{et al.} \cite{Kurby_DCASE2016_01} have proposed a method based on the late fusion of scene classification results obtained with each microphone.
Many conventional methods for multichannel ASC also apply this strategy \cite{Inoue_DCASE2018_01,Liu_DCASE2018_01}.
Imoto \textit{et al.} have proposed ASC methods using the spatial cepstrum and graph cepstrum that can be applied under an unsynchronized condition \cite{Imoto_TASLP2017_01,Imoto_IEICE2020_01}.

On the other hand, sounds recorded with smartphones or IoT devices often have missing parts caused by microphone failure, packet loss in data transmission over the network, or unreliable observations caused by clipping and wind noise.
To analyze acoustic scenes from intermittently missing observations with a single-channel microphone, Imoto and Ono have proposed a method of simultaneously analyzing acoustic scenes and estimating missing observations \cite{Imoto_TASLP2019_01}.
However, the conventional method is not for multichannel audio recordings, and the impact of partially missing channels on the ASC performance using multichannel audio recordings has not been investigated in the conventional works.

In this paper, we thus investigate the impact of partially missing channels on the performance of multichannel ASC, especially for the distributed microphone array.
In machine-learning-based multichannel ASC, missing channels cause not only losses of time-frequency and spatial information on sound sources but also a mismatch between a trained model and evaluation data.
Therefore, to realize a robust ASC system, it is important to investigate how a missing channel affects the ASC performance.
We then apply simple data augmentation methods for multichannel ASC with partially missing channels and evaluate the scene classification performance using the data augmentation methods.

The remainder of this paper is organized as follows.
In section 2, we discuss conventional acoustic scene classification using multichannel observation.
In section 3, we introduce three simple data augmentation methods for multichannel ASC with missing channels.
In section 4, we report the results of experiments carried out to evaluate the performance of ASC with partially missing channels and the impact of missing channels on the ASC performance. 
Finally, we conclude this paper in section 5.
%
%
\vspace{5pt}
\section{Conventional Methods for Scene Classification}
\label{sec:conventioinal}
Let us consider a model $f$ and model parameter ${\bi \theta}$.
The purpose of ASC is to estimate an acoustic scene label $\hat{z}$ in an evaluated sound as

\vspace{-2pt}
\begin{align}
\hat{z} = \argmax_{z} f({\bf X}, {\bi \theta}),
\end{align}
\vspace{-6pt}

\noindent where $z$ and ${\bf X} \hspace{0.7pt} ( \hspace{0.7pt} \in \hspace{-3pt} \mathbb{R}^{F \times T \times C})$ are the acoustic scene class and acoustic feature, respectively.
$F$, $T$, and $C$ are the numbers of frequency bins, time frames, and channels, respectively.
The model parameter ${\bi \theta}$ is preliminarily determined using the training dataset $\mathcal D = \{ ({\bf X}_{1}, z_{1}), ..., ({\bf X}_{l}, z_{l}), ..., $ $({\bf X}_{L}, z_{L}) \}$.
Here, ${\bf X}_{l}$ is the acoustic feature of the $l$th sound clip and $z_{l}$ indicates an acoustic scene label in the $l$th sound clip.
For the acoustic feature ${\bf X}_{l}$, the mel-band energy and MFCCs are often used.
As the model $f$, GMMs, a CNN, a ResNet-based, or a VGG-based method has often been applied.
In the neural-network-based methods, the model parameter ${\bi \theta}$ is estimated using the softmax cross-entropy loss function and the backpropagation technique.

Most conventional methods assume that there is no missing channel in a multichannel observation.
However, in the scenario of a distributed microphone array, we may have partially missing channels caused by microphone failure, in which some acoustic feature ${\bf X}_{c}$ in the $c$\hspace{0.6pt}th channel cannot be utilized in the evaluation data.
%
%
%
\vspace{1pt}
\section{Data Augmentation for Multichannel Scene Classification}
\label{sec:conventioinal}
\vspace{1pt}
In this work, we apply three data augmentation methods for multichannel scene classification with partially missing channels.
These data augmentation methods are reasonably simple to implement and enable us to investigate how partially missing channels affect the ASC performance.
%
\subsection{Channel Mask}
\label{ssec:ChMask}
The data missing in the evaluation stage causes a mismatch between the trained model and evaluation data.
To avoid this mismatch, we apply simple binary masking throughout the input time-frequency features for the random channels in the model training stage as follows:

\vspace{-2pt}
\begin{align}
{\bf X}_{l,c} = O,
\label{eq:ChMask}
\end{align}
\vspace{-6pt}

\noindent where ${\bf X}_{l,c}$ is the acoustic feature of the $l$th sound clip in the $c$\hspace{0.6pt}th channel.
$O$ is the zero matrix when the acoustic feature is the linear spectrum, whereas it is the matrix that has negative infinity values in its element when the acoustic feature is the log spectrum.
%
%
%
\subsection{Channel Overwrite and Random Copy}
\label{ssec:ChOverwrite}
When applying \textit{channel mask}, a large gap may remain between the unmasked channel in the training data and the missing channel in the evaluation data.
To bridge this gap, we apply a data augmentation method using \textit{channel overwrite} in the model training stage and a random copy in the evaluation stage.
\textit{Channel overwrite} mandatorily overwrites the time-frequency features between channels in model training as follows:

\vspace{-2pt}
\begin{align}
\hspace{1pt} {\bf X}_{l,c} = {\bf X}_{l,c'}.
\label{eq:ChOverwrite}
\end{align}
\vspace{-6pt}

\noindent In the evaluation stage, we randomly copy the acoustic features from non-missing channels to missing channels.
\begin{table}[!t]
\small
\begin{center}
\renewcommand{\arraystretch}{1.02}
\caption{Number of Recorded Audio Segments}
\vspace{-5pt}
\label{tab:audiosegments}
\vspace{0pt}
\begin{tabular}{cr}
\wcline{1-2}
\ \\[-8.0pt]
\textbf{Acoustic scene} & \textbf{\# segments} \\[-0.4pt]
\wcline{1-2}
\ \\[-8.0pt]
Absence&4,715\ \ \ \ \\[0pt]
Cooking&1,281\ \ \ \ \\[0pt]
Dishwashing&356\ \ \ \ \\[0pt]
Eating&577\ \ \ \ \\[0pt]
Other&515\ \ \ \ \\[0pt]
Social activity&1,236\ \ \ \ \\[0pt]
Vacuum cleaning&243\ \ \ \ \\[0pt]
Watching TV&4,662\ \ \ \ \\[0pt]
Working&4,661\ \ \ \ \\
\wcline{1-2}
\ \\[-8.0pt]
Total&18,246\ \ \ \ \\[-1pt]
\wcline{1-2}
\end{tabular}
\end{center}
%
\vspace{15pt}
%
\vspace{-1pt}
\footnotesize
\caption{Detailed network architecture used for
\protect\linebreak evaluation experiments}
\vspace{-5pt}
\label{tbl:parameter}
\centering
\begin{tabular}{ccc}
\wcline{1-3}
&\\[-7.5pt]
\! \textbf{Layer} \!\!\!&\!\!\! \textbf{Input size} \!\!\!\!&\!\!\!\! \textbf{Output size} \!\!\!\\
\wcline{1-3}
\!&\\[-7.2pt]
\!Conv. (7$\times$1$\times$64) + BN + ReLU \!\!\!&\!\!\! 40$\times$501$\times$16 \!\!\!\!&\!\!\!\! 40$\times$501$\times$64 \!\!\!\!\\[0pt]
\!Max pooling (4$\times$1) + Dropout (rate\hspace{1pt}=\hspace{1pt}0.2) \!\!\!&\!\!\! 40$\times$501$\times$64 \!\!\!\!&\!\!\!\! 10$\times$501$\times$64 \!\!\!\!\\[0pt]
\!Conv. (10$\times$1$\times$128) + BN + ReLU \!\!\!&\!\!\! 10$\times$501$\times$64 \!\!\!\!&\!\!\!\! 10$\times$501$\times$128 \!\!\!\!\\[0pt]
\!Conv. (1$\times$7$\times$256) + BN + ReLU \!\!\!&\!\!\! 10$\times$501$\times$128 \!\!\!\!&\!\!\!\! 10$\times$501$\times$256 \!\!\!\!\\[0pt]
\!Global max pooling + Dropout (rate\hspace{1pt}=\hspace{1pt}0.5) \!\!\!&\!\!\! 10$\times$501$\times$256 \!\!\!\!&\!\!\!\! 256 \!\!\!\!\\[0pt]
\!Dense \!\!\!&\!\!\! 256 \!\!\!\!&\!\!\!\! 128 \!\!\\[0pt]
\!Softmax \!\!\!&\!\!\! 128 \!\!\!\!&\!\!\!\! 9 \!\!\\[0pt]
\wcline{1-3}
\end{tabular}
\vspace{0pt}
%
\vspace{15pt}
%
\footnotesize
\caption{Scene classification performance with missing channels in evaluation dataset}
\vspace{-5pt}
\label{tbl:degradation}
\centering
\begin{tabular}{ccccccc}
\wcline{1-7}
&\\[-7.5pt]
\!\!\!\!&\!\!\!\! \textbf{w/o missing} \!\!\!\!&\!\!\! \textbf{1ch} \!\!\!&\!\!\! \textbf{2ch} \!\!\!&\!\!\! \textbf{4ch} \!\!\!&\!\!\! \textbf{8ch} \!\!\!&\!\!\! \textbf{12ch}\!\!\\
\wcline{1-7}
\!\!\!&\!\!\\[-7.2pt]
\!\!\textbf{Micro-Fscore}\!\!\!\!\!&\!\!\!\!\!\!96.80\%\!\!\!\!\!&\!\!\!85.50\%\!\!\!&\!\!\!74.88\%\!\!\!&\!\!\!60.93\%\!\!\!&\!\!\!38.63\%\!\!\!&\!\!\!17.67\%\!\!\\[0pt]
\!\!\textbf{Macro-Fscore}\!\!\!\!\!&\!\!\!\!\!\!93.60\%\!\!\!\!\!&\!\!\!65.90\%\!\!\!&\!\!\!45.66\%\!\!\!&\!\!\!31.44\%\!\!\!&\!\!\!13.34\%\!\!\!&\!\!\!13.98\%\!\!\\[0pt]
\wcline{1-7}
\end{tabular}
%
\vspace{19pt}
%
\footnotesize
\caption{Scene classification performance with same missing channels in training and evaluation datasets}
\vspace{-5pt}
\label{tbl:same}
\centering
\begin{tabular}{cccccc}
\wcline{1-6}
&\\[-7.5pt]
& \textbf{1ch} & \textbf{2ch} & \textbf{4ch} & \textbf{8ch} & \textbf{12ch} \\
\wcline{1-6}
&\\[-7.2pt]
\textbf{Micro-Fscore}&95.67\%&95.05\%&95.23\%&93.55\%&90.35\%\\[0pt]
\textbf{Macro-Fscore}&91.26\%&89.43\%&89.51\%&85.58\%&79.53\%\\[0pt]
\wcline{1-6}
\end{tabular}
\vspace{10pt}
\end{table}
\begin{table*}[t]
\vspace{1pt}
\small
\caption{Micro-Fscore for classification performance with data augmentation.
\protect\linebreak ``$n$ch missing'' denotes the number of missing channels in the evaluation dataset.}
\vspace{-5pt}
\label{tbl:microFscore_aug}
\centering
\begin{tabular}{lcccccc}
\wcline{1-7}
&\\[-7.5pt]
& \textbf{w/o missing} & \textbf{1ch missing} & \textbf{2ch missing} & \textbf{4ch missing} & \textbf{8ch missing} & \textbf{12ch missing} \\
\wcline{1-7}
&\\[-7.2pt]
w/o augmentation & \textbf{96.80\%} & 85.50\% & 74.88\% & 60.93\% & 38.63\% & 17.67\%\\[1pt]
\textit{Channel mask} & 95.54\% & 95.59\% & 95.45\% & 94.85\% & 92.47\% & 76.01\%\\[1pt]
\textit{Channel overwrite} + Random copy & 95.81\% & 95.78\% & 95.68\% & 95.36\%& 93.91\% & 91.43\%\\[1pt]
\textit{Channel swap} + Random copy & 95.85\% & \textbf{95.82\%} & \textbf{95.72\%} & \textbf{95.39\%} & \textbf{94.06\%} & \textbf{91.46\%}\\[0pt]
\wcline{1-7}
\end{tabular}
\vspace{15pt}
\caption{Macro Fscore for classification performance with data augmentation
\protect\linebreak ``$n$ch missing'' denotes the number of missing channels in the evaluation dataset.}
\vspace{-5pt}
\label{tbl:macroFscore_aug}
\centering
\begin{tabular}{lcccccc}
\wcline{1-7}
&\\[-7.5pt]
& \textbf{w/o missing} & \textbf{1ch missing} & \textbf{2ch missing} & \textbf{4ch missing} & \textbf{8ch missing} & \textbf{12ch missing} \\
\wcline{1-7}
&\\[-7.2pt]
w/o augmentation & \textbf{93.60\%} & 65.90\% & 45.66\% & 31.44\% & 13.34\% & 13.98\%\\[1pt]
\textit{Channel mask} & 90.63\% & 90.76\% & 90.42\% & 88.81\% & 84.28\% & 51.99\%\\[1pt]
\textit{Channel overwrite} + Random copy & 90.75\% & 90.75\% & 90.53\% & 90.06\%& 88.21\% & 83.98\%\\[1pt]
\textit{Channel swap} + Random copy & 90.74\% & \textbf{90.75\%} & \textbf{90.54\%} & \textbf{90.13\%} & \textbf{88.27\%} & \textbf{84.07\%}\\[0pt]
\wcline{1-7}
\end{tabular}
\vspace{10pt}
\end{table*}
%
%
%
\subsection{Channel Swap and Random Copy}
\label{ssec:ChSwap}
\textit{Channel mask} and \textit{channel overwrite} lose time-frequency information since we discard time-frequency features in the training stage.
To train the scene classification model without wasting time-frequency information, we apply a data augmentation method using \textit{channel swap}.
\textit{Channel swap} simply swaps the time-frequency features between channels in model training as follows:

\vspace{-3pt}
\begin{align}
\begin{cases}
\hspace{1pt} {\bf X}_{l,c} = {\bf X}_{l,c'}\\
\hspace{1pt} {\bf X}_{l,c'} = {\bf X}_{l,c}.
\end{cases}
\label{eq:ChSwap}
\end{align}
\vspace{-1pt}

\noindent In the evaluation stage, we randomly copy the acoustic features from non-missing channels to missing channels as with the data augmentation in \textit{channel overwrite}.
%
%
%
%
\vspace{2pt}
\section{Experiments}
\label{sec:experiments}
\vspace{1pt}
\subsection{Experimental Conditions}
\label{ssec:conditions}
We evaluate the impact of a missing channel on the performance of scene classification using various data augmentation methods.
To evaluate the performance, we use the development dataset of DCASE2018 Challenge Task 5 \cite{Dekkers_DCASE2018_01}, which is a derivative of the SINS dataset \cite{Dekkers_DCASE2017_01}.
We construct each 10 s audio segment with sounds recorded by four microphone arrays, each of which consists of four linearly arranged microphones; that is, each audio segment contains 16 channels.
As shown in Table~\ref{tab:audiosegments}, the dataset contains 18,246 audio segments, and we split them into the same 4-fold cross-validation setup as in DCASE2018 Challenge Task 5.

For the acoustic features, we use the 40-dimensional log mel-band energy, which has a frame length of 40 ms with hop size  of a 20 ms.
In this paper, we regard the missing channels as silent with zeros filled in the time domain.
As the classification model, we apply the same network proposed by Inoue {\it et al.} \cite{Inoue_DCASE2018_01}, which achieved the best score in DCASE2018 Challenge Task 5, except for the input channel size of the network.
The detailed network structure is shown in Table~\ref{tbl:parameter}.
We utilize the RAdam optimizer \cite{Liu_ICLR2020_01} with a learning rate of 0.001.
For each method, we conduct the evaluation experiment 16 (random combinations of missing channels) $\times$ 4 (fold) times.
\begin{figure*}[t]
\vspace{15pt}
\begin{tabular}{ccc}
\begin{minipage}[t]{0.31\hsize}
\centering
\includegraphics[height=1.09\textwidth]{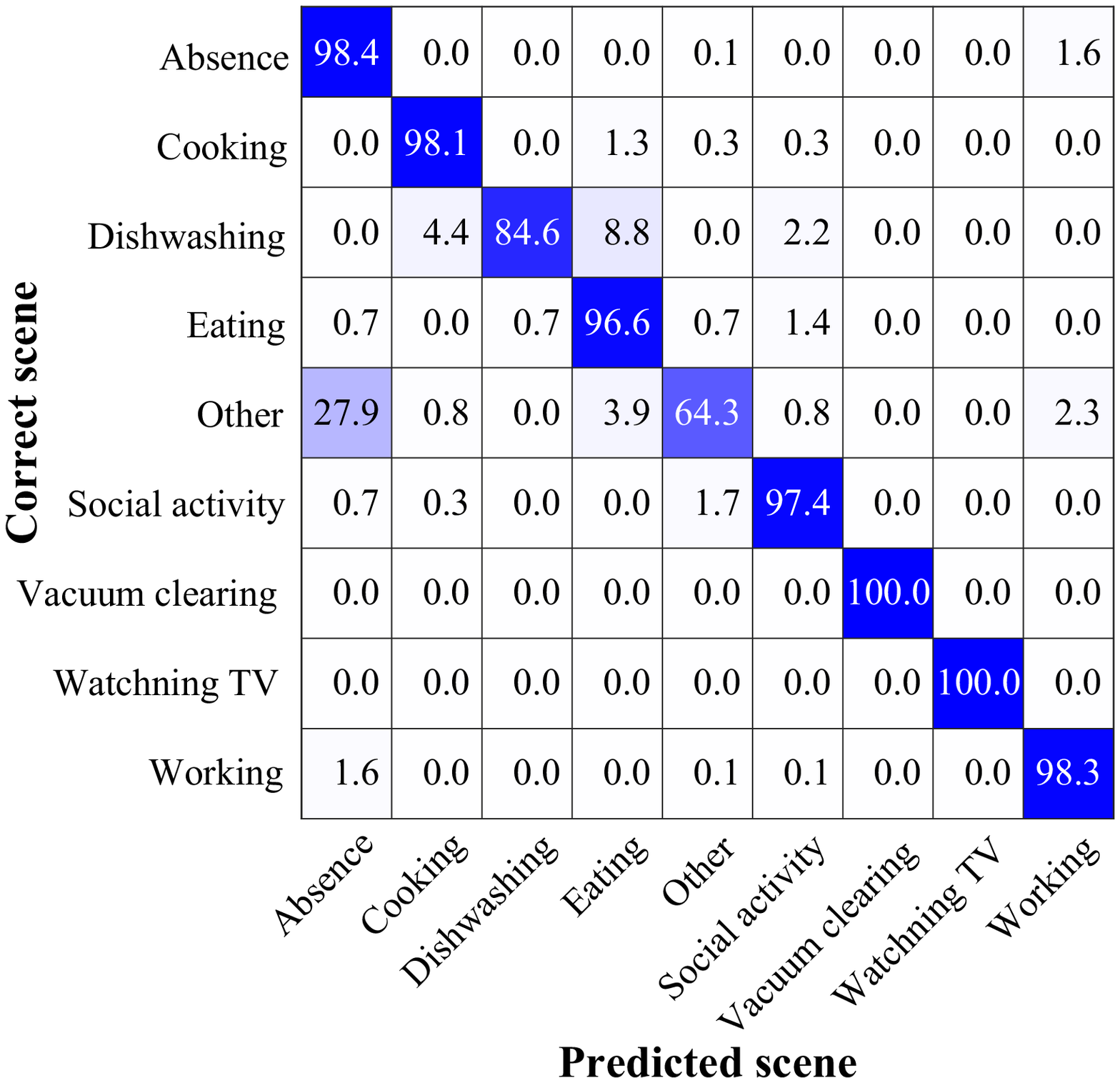}
\vspace{-14pt}
\caption{Example of scene classification result without missing channels (recall, \%)}
\label{fig:nomiss}
\end{minipage}
\hspace*{9pt}
\begin{minipage}[t]{0.31\hsize}
\centering
\hspace*{12pt}
\includegraphics[height=1.09\textwidth]{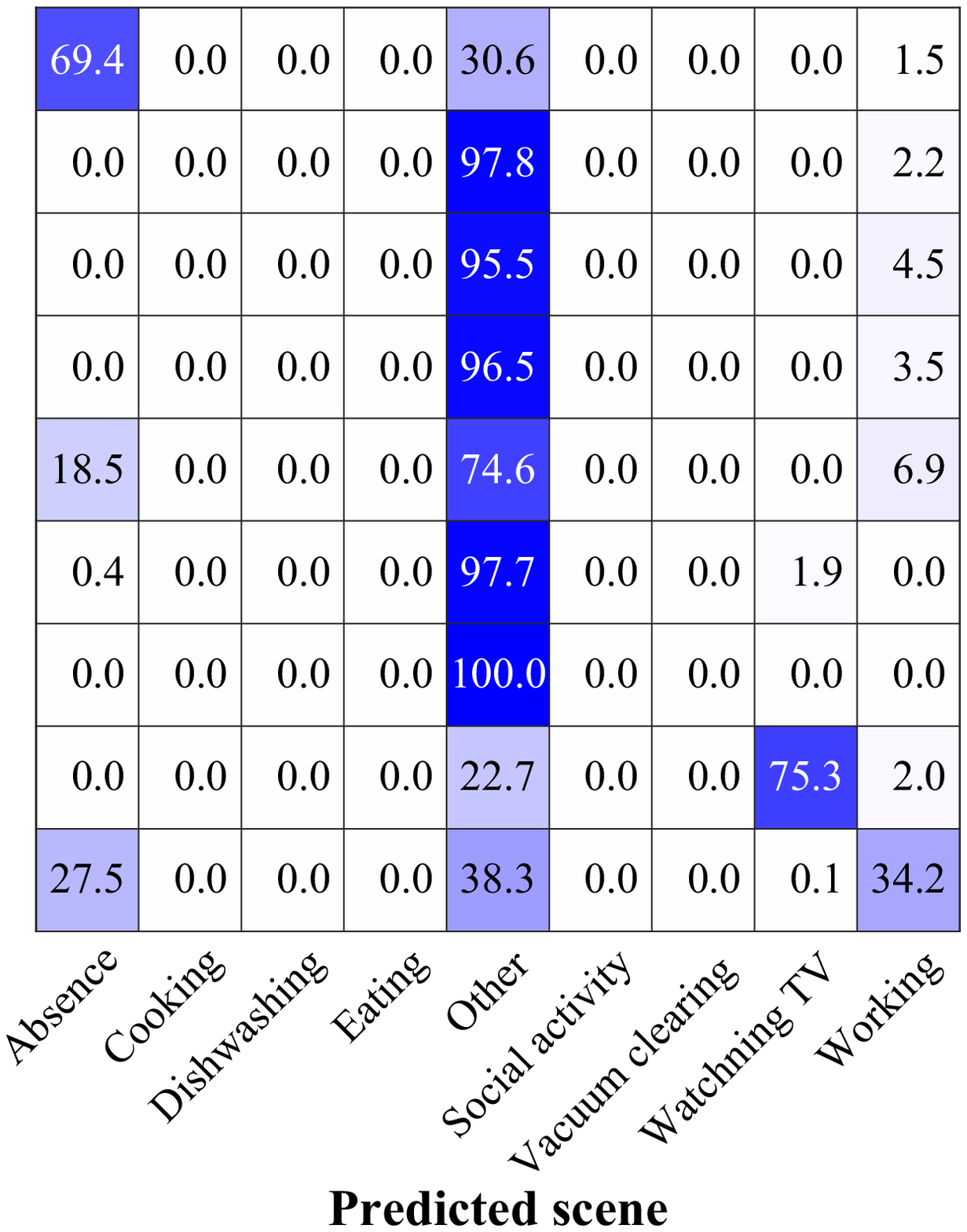}
\vspace{-4pt}
\caption{Example of scene classification result with four missing channels in evaluation data (recall, \%)}
\label{fig:missing}
\end{minipage}
\hspace*{7pt}
\begin{minipage}[t]{0.31\hsize}
\centering
\hspace*{-3pt}
\includegraphics[height=1.09\textwidth]{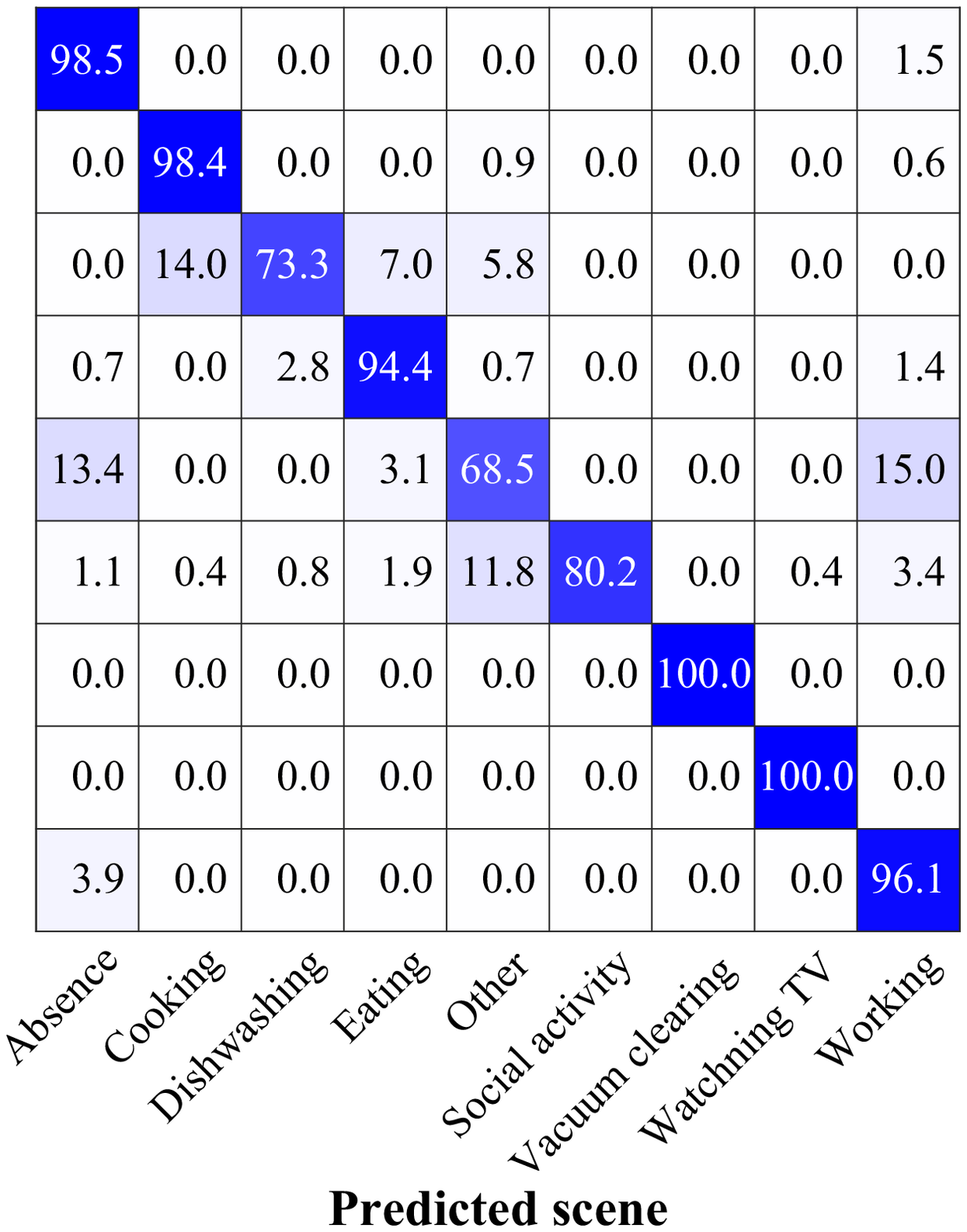}
\vspace{-4pt}
\caption{Example of scene classification result based on \textit{channel swap} with four missing channels in evaluation data (recall, \%)}
\label{fig:ChSwap}
\end{minipage}
\end{tabular}
\vspace{10pt}
\end{figure*}
%
%
\subsection{Experimental Results}
\label{ssec:results}
\subsubsection{Impact of Missing Channel on Classification Performance}
We evaluate the performance degradation caused by missing channels in the evaluation data.
Table~\ref{tbl:degradation} shows the scene classification performance in terms of micro- and macro-Fscore.
The result shows that the missing channels cause severe performance degradation in multichannel ASC.

To investigate how the missing channels affect the ASC performance, we also evaluate the ASC performance with the same channels missing in the training and evaluation datasets.
Table~\ref{tbl:same} shows the scene classification performance in terms of micro- and macro-Fscore.
The results show that the performance degradation is significantly smaller than the corresponding results in Table~\ref{tbl:degradation}, even though some channels are missing in the training dataset.
This indicates that, in multichannel ASC, the mismatch between the trained model and evaluation data is a much more severe problem than missing spectral and spatial information.
Thus, the mismatch between the trained model and evaluation data must be addressed preferentially in multichannel ASC with partially missing channels.
%
%
\vspace{8pt}
\subsubsection{Evaluation of Data Augmentation Technique for Multichannel ASC}
\vspace{2pt}
We next evaluate the ASC performance with the proposed data augmentation methods.
In this experiment, we randomly select a number of channels from 0 to 8 for data augmentation in each iteration of model training.
Tables~\ref{tbl:microFscore_aug} and \ref{tbl:macroFscore_aug} show the scene classification performance in terms of micro- and macro-Fscore with the proposed data augmentation methods.
The results show that the three data augmentation methods achieve reasonable performances.
In particular, \textit{channel overwrite} and \textit{channel swap} achieve comparable ASC performance to the result without missing channels.
Comparing these results with Table~\ref{tbl:same} indicates that \textit{channel overwrite} and \textit{channel swap} can almost completely avoid the mismatch between the trained model and evaluation data.
%
%
%
%
%
\vspace{8pt}
\subsubsection{Datailed Scene Classification Performance}
\vspace{2pt}
Figs.~\ref{fig:nomiss}--\ref{fig:ChSwap} show the detailed scene classification performance using no data augmentation method and \textit{channel swap} with four channels missing.
The results show that most of the audio segments are predicted as ``other.''
On the other hand, the classification result using \textit{channel swap} achieves comparable performance to that without missing channels.
From these results, we conclude that, in multichannel ASC, partially missing channels may cause a severe degradation of the ASC performance, and avoiding the mismatch between the trained model and evaluation data is important to achieve a robust ASC system in a realistic situation.
%
%
%
\section{Conclusion}
\label{sec:conclusion}
In this paper, we investigated the impact of partially missing channels on the ASC performance using multichannel audio recordings obtained using a distributed microphone array.
We also proposed three data augmentation methods for multichannel ASC: \textit{channel mask}, \textit{channel overwrite}, and \textit{channel swap}.
The experimental results showed that, in multichannel ASC, the mismatch between the trained model and evaluation data is a much more severe problem than missing spectral and spatial information.
To avoid this negative impact on the ASC performance, the data augmentation based on \textit{channel overwrite} and \textit{channel swap} is effective and can avoid the performance degradation caused by the model mismatch.
%
%
\section*{Acknowledgment}
This work was supported by JSPS KAKENHI Grant Number JP20H00613, JP19K20304, and KDDI Foundation.
%
%
\vspace{3pt}
\bibliographystyle{IEEEtran}
\bibliography{IEEEabrv,KeisukeImoto11,EUSIPCO2021refs}
\end{document}